\documentclass[floats,floatfix,showpacs,amssymb,prd,twocolumn,superscriptaddress,nofootinbib,nolongbibliography,reprint,preprintnumbers]{revtex4-1}

\usepackage{amssymb,amsmath,verbatim,mathtools,needspace,enumitem,etoolbox,graphicx,physics,microtype,afterpage,xspace,tabularx,lmodern,multirow,braket,booktabs}
\usepackage{gensymb}
\usepackage[normalem]{ulem}
\usepackage[dvipsnames, usenames]{xcolor}
\definecolor{linkcolor}{rgb}{0.0,0.3,0.5}
\usepackage[unicode, colorlinks=true, linkcolor=linkcolor, citecolor=linkcolor, filecolor=linkcolor, urlcolor=linkcolor, linktocpage, breaklinks]{hyperref}
\usepackage[all]{hypcap}
\usepackage{subfigure}
\usepackage[T1]{fontenc}
\usepackage[utf8]{inputenc}
\usepackage[usenames,dvipsnames]{xcolor}
\hypersetup{colorlinks=true,citecolor=romared,linkcolor=romared,urlcolor=romared}

\definecolor{romared}{RGB}{142,0,28}

\newcommand{\be}{\begin{equation}}
\newcommand{\ee}{\end{equation}}

\def\be{\begin{equation}}
\def\ee{\end{equation}}
\newcommand{\beq}{\begin{eqnarray}}
\newcommand{\eeq}{\end{eqnarray}}

\usepackage{aas_macros}
\usepackage{makecell}
\usepackage{soul}
\usepackage{amssymb}
\usepackage{longtable}

\usepackage{lipsum}
\usepackage{orcidlink}

\newcolumntype{Y}{>{\centering\arraybackslash}X}

\begin{document}

\title{Excitation region of Kerr black hole quasinormal modes from Stokes geometry}

\author{Naritaka Oshita
\orcidlink{0000-0002-8799-1382}}
\email{oshita@phys.kindai.ac.jp}
\affiliation{Department of Physics, Kindai University, Osaka 577-8502, Japan}
\affiliation{RIKEN iTHEMS, Wako, Saitama, 351-0198, Japan}

\preprint{RIKEN-iTHEMS-Report-26}

\begin{abstract}
We investigate the excitation region of black hole quasinormal modes (QNMs) from both analytical and numerical perspectives. 
On the analytical side, we propose that the QNM excitation radius is identified by the dominance switching of WKB solutions across an anti-Stokes line. 
Based on this picture, we derive the condition for the QNM excitation radius in Kerr spacetime. 
In the Schwarzschild limit with mass $M$, we reproduce the previously known value $r=2.556929 M$, which differs from the light ring radius $r=3M$. 
We also show that the excitation radius is independent of the angular mode $\ell$ in the high-overtone limit. 
As an independent approach, we employ the numerical convergence test of QNM-plus-tail expansion and obtain values consistent with the Stokes geometry in the high-overtone limit at low and intermediate spins. 
In the extremal limit, the QNM convergence radius approaches the outer horizon, which is captured by the Stokes geometry of the zero-damping modes, rather than the high-overtone limit. 
This is consistent with the fact that the ringdown is dominated by zero-damping modes in the extremal limit. 
Based on the complementary analysis of Stokes geometry and the convergence test, we argue that in general, the QNM excitation radius depends on the QNM overtone number, giving rise to an effective QNM convergence region.
\end{abstract}

\maketitle

%%%%%%%%%%%%%%%%%%%%%%%%
\section{Introduction}
%%%%%%%%%%%%%%%%%%%%%%%%

Black hole (BH) spectroscopy \cite{Kokkotas:1999bd,Berti:2009kk,Berti:2025hly} provides a powerful approach to probing both the strong-gravity region near BHs and the underlying theory of gravity. 
In this framework, the ringdown waveform emitted by a perturbed BH is modeled as a superposition of BH quasinormal modes (QNMs), and the BH parameters are inferred from their characteristic frequencies and damping rates \cite{LIGOScientific:2026wpt,LIGOScientific:2025rid,LIGOScientific:2025wao}.

However, it remains unclear from what time onward the QNM description becomes valid for BH merger waveforms, a longstanding issue often referred to as the starting-time problem. 
Motivated by this, in this work we investigate the location at which QNMs are excited, which we refer to as the QNM excitation radius.

QNM excitation factors \cite{Leaver:1986gd,Sun:1988tz,Andersson:1995zk,Glampedakis:2001js,Glampedakis:2003dn,Berti:2006wq,Zhang:2013ksa} are key quantities that characterize the excitability of QNMs and also play a role in determining the convergence of QNM expansions.
In the Schwarzschild case with mass $M$, the convergence of QNM excitation factors has been observed near the origin of the conventional tortoise coordinate ($x \sim 0$) \cite{Andersson:1996cm}, indicating the existence of a characteristic location associated with QNM excitation.
The recent detailed studies \cite{DeAmicis:2025xuh,Arnaudo:2025uos,Arnaudo:2026tcy,DeAmicis:2026wqd} identify that the QNM excitation originates at $x = 0$ or $r = 2.556929 M$. 
The convergence of Kerr QNM excitation factors has also been investigated in Ref.~\cite{Oshita:2024wgt}.
\begin{figure}[t]
\centering
\includegraphics[width=1\linewidth]{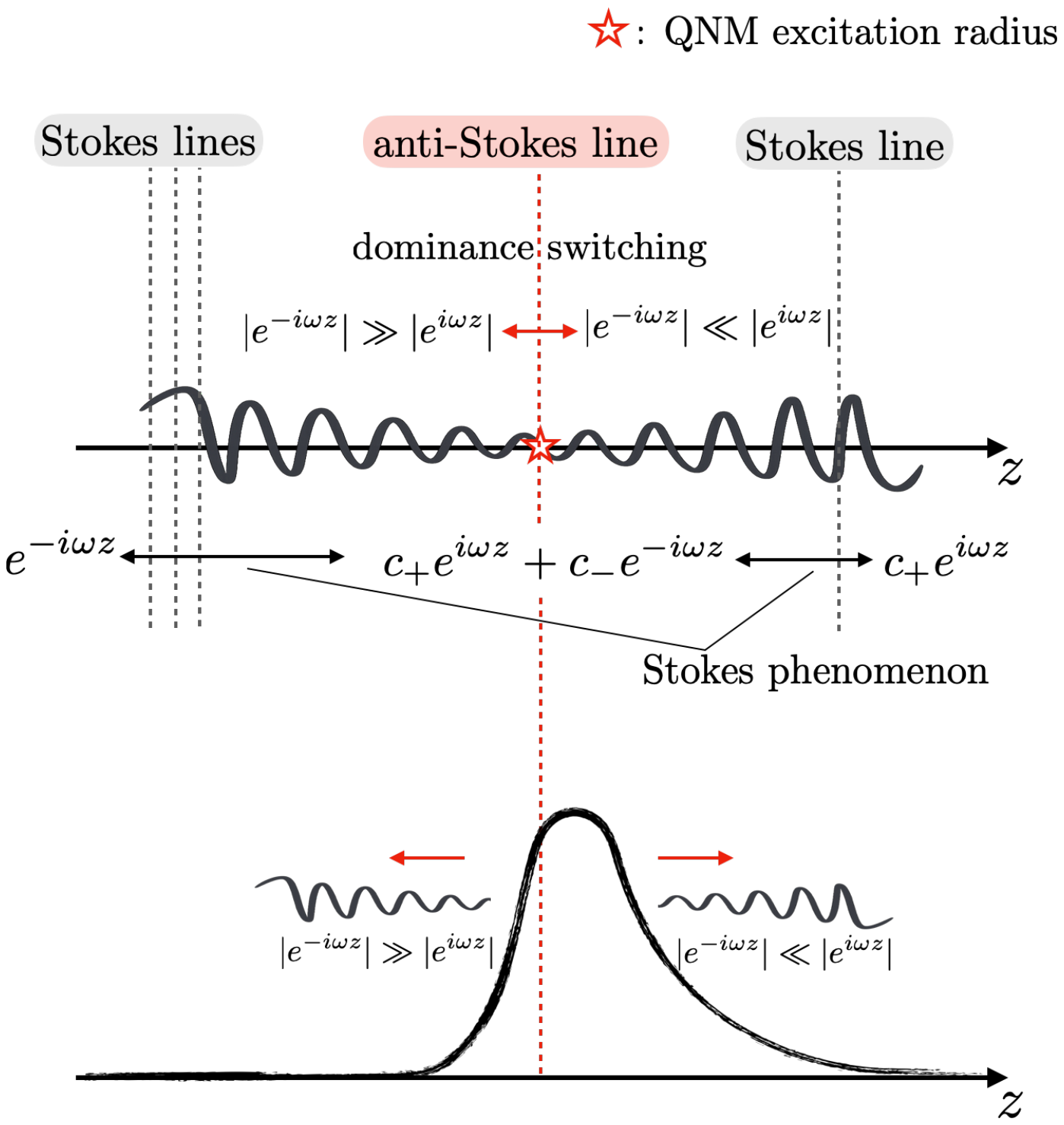}
\caption{A schematic picture describing the relation between the Stokes geometry and the QNM excitation radius.
The coordinate $z$ stands for the generic tortoise coordinate.
}
\label{fig_schematic}
\end{figure}

The BH perturbations has previously been studied from the viewpoint of the Stokes geometry, especially in the context of monodromy analysis \cite{Motl:2003cd,Andersson:2003fh,Natario:2004jd,Keshet:2007nv,Keshet:2007be} and the exact WKB analysis \cite{Miyachi:2025ptm,Miyachi:2025dyk,Hatsuda:2026ghx}.
Among these studies, Keshet et al. \cite{Keshet:2007nv,Keshet:2007be} interpreted one particular anti-Stokes line as the ``excitation line'' supporting the QNM bound state and leading to the QNM quantization condition.
On the other hand, in our work, we identify the intersection of the anti-Stokes line and the real-$r$ axis as the QNM excitation region, supported by its consistency with the convergence region of the QNM-plus-tail expansion, as explained below.

We investigate the origin of QNM excitation in Kerr spacetime, with mass $M$ and dimensionless spin parameter $\chi$, by combining two complementary approaches: an analytical approach based on the Stokes geometry and a numerical study of the convergence of the QNM and branch-cut expansions using the Mano-Suzuki-Takasugi (MST) method \cite{Mano:1996vt}.
Comparing the QNM excitation radius ($r=r_{\rm Stokes}$) defined by the Stokes geometry and the QNM convergence radius ($r=r_{\rm con}$) obtained from the QNM convergence test, we conclude that the excitation region of QNMs in Schwarzschild and Kerr spacetimes can be understood in terms of the underlying Stokes geometry.

We find that the value of $r_{\rm Stokes}$ obtained in the large-overtone regime agrees with the QNM convergence radius $r_{\rm con}$, especially for low and intermediate BH spins.
Especially, in the Schwarzschild case, we confirm that the QNM convergence radius is independent of the angular mode $\ell$: $r_{\rm con} \simeq 2.5569 M$ for $\ell = 2$, $3$ and $4$.
This is consistent with our prediction based on the Stokes geometry, which is also independent of $\ell$ in the high-overtone limit.

\begin{figure*}[t]
\centering
\includegraphics[width=0.45\textwidth]{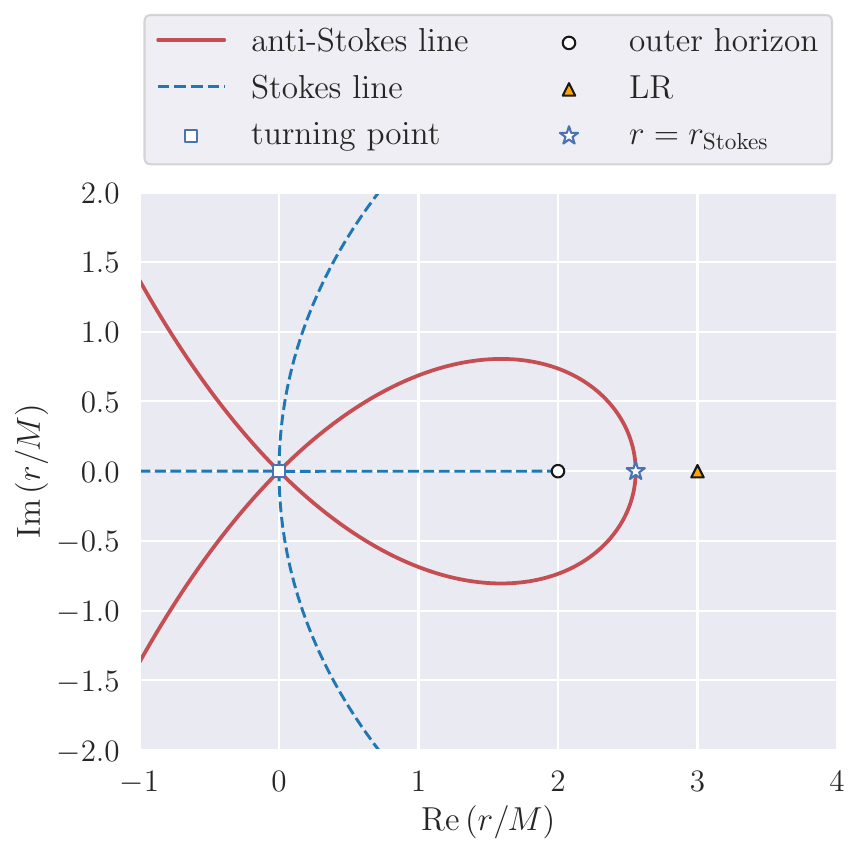}
\includegraphics[width=0.45\textwidth]{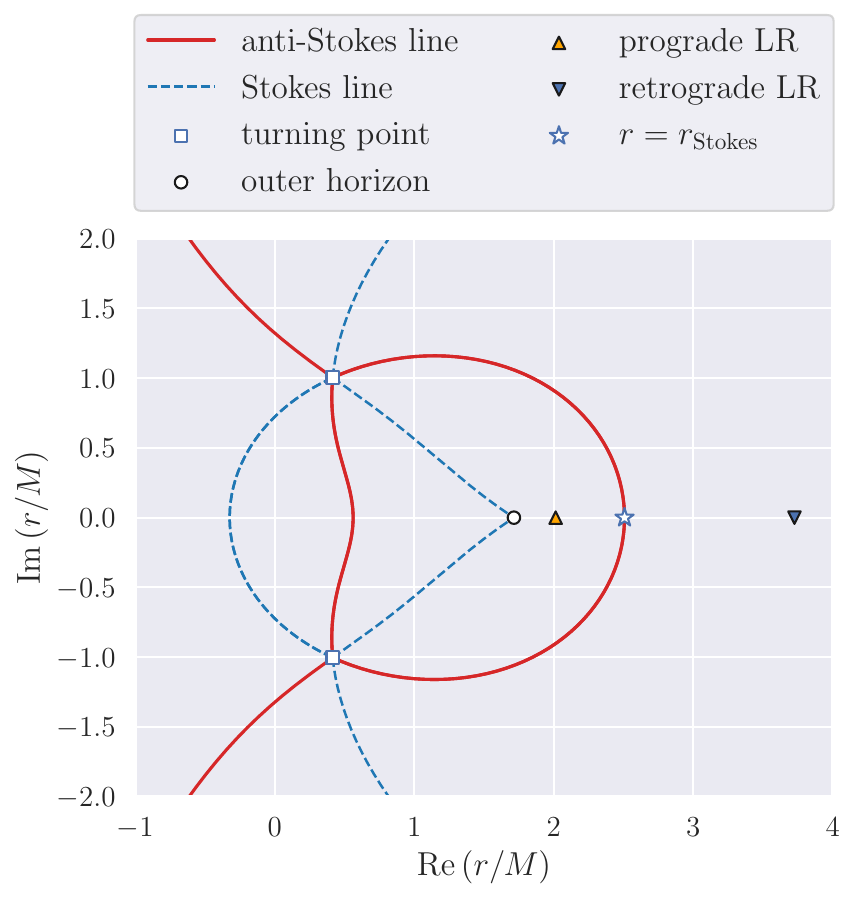}
\caption{
Stokes geometry in the Schwarzschild BH (left) and in the Kerr BH with $\chi = 0.7$ (right). We here take the limit of $\text{Im}(\omega) \to - \infty$.
Stokes (blue dashed) and anti-Stokes lines (red solid) emanate from turning points (square).
The anti-Stokes line has an intersection with the axis of $\text{Re} (r)$ at $r = r_{\rm Stokes}$ (star). 
For the case of Schwarzschild BH, we find $r_{\rm Stokes}/M = 2.5569291$, i.e., the root of the conventional tortoise coordinate $r+\log[r/(2M) -1] = 0$, which agrees with the literature \cite{DeAmicis:2025xuh,Arnaudo:2025uos,Arnaudo:2026tcy,DeAmicis:2026wqd}.
The blue dotted lines stand for the Stokes lines. The circle and triangle markers are the radius of the outer horizon and the radii of LRs, respectively.
}
\label{fig_stokes_0000_0700}
\end{figure*}
We further extend the analysis of the Stokes geometry to the near-extremal regime. 
We find that the value of $r_{\rm Stokes}$ in the large-overtone limit ($\text{Im} (\omega) \to -\infty$) differs from $r_{\rm con}$ in the near-extremal case.
As the ringdown is dominated by zero-damping modes (ZDMs) \cite{Yang:2013uba}, we compute the value of $r_{\rm Stokes}$ for ZDMs in the extremal limit, and find that it is consistent with the value of $r_{\rm con}$. 
This is likely because many QNMs become long-lived in the near-extremal regime, making the large-overtone limit less representative of the physically relevant mode spectrum. 
These findings, therefore, lead us to conclude that, in general, different overtones are excited at different radii.
Indeed, different frequencies $\omega$ give rise to different Stokes geometries and anti-Stokes lines.

This paper is organized as follows.
In Sec.~\ref{sec_stokes_geo}, we briefly review the Stokes phenomenon and the dominance switching of QNM eigenfunctions across anti-Stokes lines. We then discuss, in the high-overtone regime, how anti-Stokes lines are related to the region of QNM excitation.
We also compute the QNM excitation radius for a wide range of black hole spins in the high-overtone limit.
In Sec.~\ref{sec_excitation_radius}, we compare the obtained QNM excitation radius from the Stokes geometry with the QNM convergence radius, $r_{\rm con}$, obtained from the QNM convergence test. 
We further show that, in the near-extremal regime, the excitation radius predicted from the Stokes geometry of ZDMs in the extremal limit is consistent with the numerical results. We also discuss the dependence of the QNM excitation radius, $r_{\rm Stokes}$, on the overtone number.
Finally, we summarize our results and discuss future directions in Sec.~\ref{sec_conclusion}.
Appendix~\ref{app_spiral}, demonstrates that anti-Stokes lines develop logarithmic spirals when the QNM frequencies are large but finite, in the high-overtone regime, and discusses the physical interpretation of this behavior.
Appendix~\ref{app_methodology} provides the computational details of the QNM and branch-cut expansions, including the number of overtones used in the analysis and the evaluation of the branch-cut contribution.

We restrict our analysis to the gravitational perturbation with spin weight $s=-2$ and angular mode of $\ell = m=2$ unless otherwise stated, with a few results for higher angular modes included for comparison.

%%%%%%%%%%%%%%%%%%%%%%%%
\section{Stokes geometry and the site of QNM excitation}
\label{sec_stokes_geo}
%%%%%%%%%%%%%%%%%%%%%%%%
A fundamental question in BH spectroscopy is when the QNM description of BH merger waveforms is valid. 
In particular, in the extreme-mass-ratio regime, this issue is closely related to the region from which ringdown is radiated toward both the horizon and spatial infinity during the relaxation of the perturbations around the Teukolsky potential barrier.
This question is also closely relevant to the convergence of the QNM expansion of merger waveforms.
The convergence is controlled by the behavior of the excitation coefficients in the high-overtone limit. 
It is therefore natural to investigate the excitation of highly damped QNMs.
To this end, let us consider the Stokes geometry in the Teukolsky equation with the asymptotic limit, i.e., $\omega_{\rm R} / \omega_{\rm I} \to 0$, where $\omega_{\rm R} \coloneqq \text{Re} (\omega)$ and $\omega_{\rm I} \coloneqq \text{Im} (\omega)$.
Here we propose that the excitation region of QNMs can be identified by the intersection between an anti-Stokes line and the real-$r$ axis, as the anti-Stokes line gives the site at which the dominance of incoming and outgoing modes in the QNM eigenfunction switches.
This intersection therefore identifies the location from which the incoming and outgoing mode propagate towards the horizon and spatial infinity, respectively.

In this regime, the radial equation reduces to the form \cite{Keshet:2007nv,Keshet:2007be}
\begin{equation}
\left[
\frac{d^2}{dr^2}
+\omega^2 V^2(r)
\right] \psi(r)=0,
\label{Teuko_high_overtone}
\end{equation}
where
\begin{align}
\begin{split}
V &\equiv \frac{\sqrt{q_0 + {\cal O}(\omega^{-1}) }}{\Delta}\,,\\
q_0 &\equiv (r^2+a^2)^2 - a^2\Delta\,,
\end{split}
\label{V_q0}
\\
z &\equiv \int^r_{r_0} V(r') dr'\,,
\label{integration_general_tortoise}
\end{align}
and $\Delta \equiv r^2 -2Mr +a^2$.
The equation admits two independent WKB solutions,
\begin{equation}
\psi_{\pm}=e^{\pm i\omega z}\,,
\end{equation}
which reduces to the plane wave solution at the horizon and at infinity as $\psi_{\pm} = e^{\pm i \omega z} \sim e^{\pm i \omega x}$ with the conventional tortoise coordinate $x \coloneqq \int dr (r^2+a^2)/\Delta$.
In the WKB analysis, the end point of the integration $r_0$ in Eq.~\eqref{integration_general_tortoise} is identified as the {\it turning point}, which is the root of $V(r_0) =0$, and $r_0$ is a complex value in general.
A turning point is a key configuration as it governs the behavior of the WKB solutions $\psi_{\pm}$.
Stokes lines and anti-Stokes lines, on which $\text{Im} (i\omega z) = 0$ and $\text{Re} (i\omega z) = 0$, respectively, emanate from the turning point.

Let us consider the QNM eigenfunction from the point of view of the Stokes geometry.
The QNM eigenfunction is a superposition of the two independent WKB solutions, $\psi_{+}$ and $\psi_-$, at the intermediate zone due to the Stokes phenomenon \cite{Miyachi:2025ptm}.
In general, the WKB solution $\psi_{\pm}$ is either subdominant or dominant.
As is shown in FIG.~\ref{fig_schematic}, $\psi_+$ ($\psi_-$) is exponentially large and $\psi_-$ ($\psi_+$) is exponentially small at $r \to \infty$ ($r \to r_+$) for the QNM eigenfunction as $\text{Im}(\omega) < 0$.
The dominance of the solutions switches at the anti-Stokes line (red dashed line in FIG.~\ref{fig_schematic}).
Crossing a Stokes line, the coefficient of the subdominant mode undergoes a jump\footnote{The Stokes phenomenon should be understood as a change in the asymptotic representation of the analytic solution between neighboring Stokes regions, rather than a discontinuity of the solution itself.}, known as the Stokes phenomenon. 
If $\psi_-$ is subdominant, the Stokes phenomenon leads to 
\begin{equation}
\underbrace{c_+ \psi_{+} + c_- \psi_{-}}_{\text{intermediate zone}} \to \underbrace{c_+ \psi_{+} + (c_- +k c_+) \psi_{-}}_{\text{spatial infinity}}\,, 
\label{Stokes_phenomenon_example}
\end{equation}
where $k = +i$ ($-i$) for a counterclockwise (clockwise) crossing of the Stokes line.
At far region, Stokes lines have an intersection with the real $r$-axis.
This causes the Stokes phenomenon, by which the coefficient of $\psi_-$ vanishes, i.e., $(c_- +k c_+) = 0$ in Eq.~\eqref{Stokes_phenomenon_example}, to satisfy the outgoing boundary condition at infinity.
In other words, the QNM eigenfunction is a linear combination of $\psi_-$ and $\psi_+$ at the intermediate region.
As $z(r) \sim \log (r-r_+)$ around the outer horizon, the Stokes lines near the horizon have the logarithmic spiral \cite{Miyachi:2025ptm,Miyachi:2025dyk}, which also causes Stokes phenomenon near the horizon, making the solution purely ingoing mode, $\psi_-$, in the horizon limit.
The anti-Stokes line intersects the real-$r$ axis at $r=r_{\rm Stokes}$ (star marker in FIG.~\ref{fig_schematic}), which typically lies in the vicinity of the effective potential barrier as is shown later. 
At the intersection, the dominance between the ingoing and outgoing modes switches. 
This suggests a physical picture in which the QNM excitation originates at $r=r_{\rm Stokes}$ and dissipates toward both the horizon and spatial infinity. 
Motivated by this interpretation, we identify $r=r_{\rm Stokes}$ as the location where QNMs are effectively excited. 
Throughout this paper, we refer to this radius as the \emph{QNM excitation radius}, or simply the \emph{excitation radius}.

FIG.~\ref{fig_stokes_0000_0700} shows the Stokes geometry of the Teukolsky equation in the high-overtone (or high-frequency) limit \eqref{Teuko_high_overtone} with $\chi = 0$ and $\chi = 0.7$.
We tabulate the excitation radius in the high-overtone limit for various spins in Table~\ref{tab:stokes_radius}.
\begin{table*}[t]
\centering
\caption{QNM excitation radius $r_{\rm Stokes}$ with respect to the BH spin $\chi$ in the high-overtone limit ($\omega_{\rm I} \to -\infty$).
In the extremal limit, $\chi \to 1$, the excitation radius is also computed for the ZDM $\omega = m \Omega_{\rm H}$ (bracketed value).
The values of $r_{\rm con}$ defined with $\epsilon_{\rm th} = 10^{-2}$ is shown. 
The angular modes are set to $(\ell,m) = (2,2)$. 
The values of $r_{\rm con}$ with $\ell =3$ and $4$ are also shown for the Schwarzschild case (bracketed values).
}
\label{tab:stokes_radius}
\begin{tabular}{ccc}
\hline\hline
$\chi$ & $r_{\rm Stokes}/M$
& $r_{\rm con} / M \ (\epsilon_{\rm th} = 10^{-2})$ \\
\hline
0 &    2.5569291 & 2.557 [2.551, 2.551]\footnote{The bracketed values are $r_{\rm Stokes}/M$ for $\ell = 3$ and
$\ell =4$, respectively.} \\
0.1 &  2.5601873 & 2.518 \\
0.2 &  2.5617647 & 2.504 \\
0.3 &  2.5597783 & 2.486 \\
0.4 &  2.5535950 & 2.471 \\
0.5 &  2.5428780 & 2.450 \\
0.6 &  2.5274019 & 2.419 \\
0.7 &  2.5069827 & 2.368 \\
0.8 &  2.4814432 & 2.311 \\
0.9 &  2.4505929 & 2.263 \\
0.99 & 2.4181073 & 2.055 \\
0.999 &2.4146055&  1.702 \\
0.99999&2.4142175 & 1.251 \\
$\chi \to 1$ & 2.4142136 [1.1051643]\footnote{The bracketed value is $r_{\rm Stokes}/M$ for the ZDM in the extremal limit, i.e., $\omega = m \Omega_{\rm H}$.}& --- \\
\hline\hline
\end{tabular}
\end{table*}
In the Schwarzschild case, we find $r_{\rm Stokes}/M = 2.5569291$, which matches with the root of $x_{\rm Sch} (r) \coloneqq r + 2M \log\left[ r/(2M)  -1 \right] = 0$ \cite{DeAmicis:2025xuh,Arnaudo:2025uos,Arnaudo:2026tcy,DeAmicis:2026wqd}.
This is not a coincidence.
From the anti-Stokes-line condition $\text{Re}(i \omega z) = 0$ with the high-overtone approximation $\omega_{\rm I} \to - \infty$, we have
\begin{equation}
    \text{Re}(z) = 0\,.
    \label{Sch_anti_stokes_cond}
\end{equation}
For $\chi = 0$, it leads to
\begin{equation}
    x_{\rm Sch} (r_{\rm Stokes}) = 0\,,
    \label{condition_SCH_stokes_ind_ell}
\end{equation}
since
\begin{equation}
    z = \int_{r_0}^r V(r) = \underbrace{r + 2M \log \left[ \frac{r}{2M} -1 \right]}_{= x_{\rm Sch} (r)} - 2i \pi M\,,
\end{equation}
where the second term, which is pure-imaginary, has no contribution in the condition of anti-Stokes line \eqref{Sch_anti_stokes_cond}.
Note that the condition for $r_{\rm Stokes}$ in the high-overtone limit \eqref{condition_SCH_stokes_ind_ell} is independent of the angular mode $\ell$, which will be confirmed by an independent numerical approach, i.e., the QNM convergence test, in the next section.

In the same manner, we can derive the condition satisfied by the excitation radius, $r=r_{\rm Stokes}$ $\in {\mathbb R}$, in the high-overtone limit for general spins:
\begin{equation}
    \text{Re} \left( \int_{r_0^{\pm}}^r \frac{\sqrt{q_0 (r_{\rm Stokes})}}{\Delta (r_{\rm Stokes})} \right) = 0\,,
\end{equation}
where
\begin{equation}
    r_0^{\pm} = \left( \frac{1\pm i \sqrt{3}}{2 \times 3^{1/3}} \frac{\chi^2}{\Xi} - \frac{1 \mp i \sqrt{3}}{2 \times 3^{2/3}} \Xi \right) M\,,
\end{equation}
and $\Xi \coloneqq \left\{ 9 \chi^2 \left[ \sqrt{1+\chi^2/27} -1 \right] \right\}^{1/3}$.
In either case of the end point, $r_0^{+}$ or $r_0^-$, we obtain the same value of $r_{\rm Stokes}$.
The values of $r_{\rm Stokes}$ with respect to $\chi$ are tabulated in Table~\ref{tab:stokes_radius}.
It is important to note that the prograde and retrograde light rings (LRs) have different radii, denoted by $r_{{\rm LR}+}$ and $r_{{\rm LR}-}$, respectively, for non-zero spins:
\begin{equation}
r_{{\rm LR}\pm} \coloneqq 2 M \left\{ 1+\cos \left[ \frac{2}{3} \arccos{(\mp \chi)} \right] \right\}\,.
\end{equation}
As such, it is highly nontrivial where the QNM excitation happens for BHs with non-zero spins.

We emphasize that the present analysis is restricted to the asymptotic limit $\omega_{\rm I} \to -\infty$, corresponding to the Stokes geometry with $|\omega_R/\omega_I| \to 0$. 
For a large but finite QNM frequency with $0<|\omega_R/\omega_I|\ll1$, the anti-Stokes lines acquire a spiral structure, leading to a sequence of intersections with the real-$r$ axis clustered around the asymptotic excitation radius (see Appendix~\ref{app_spiral}). 
More generally, the Stokes geometry, and hence the excitation radius, depends on the QNM frequency. 
Consequently, the excitation radius is generally overtone dependent. 
As we will demonstrate in the next section, this dependence becomes especially significant in the near-extremal regime.

\section{Excitation radius and the convergence test of excitation factors}
\label{sec_excitation_radius}

Here, we examine the convergence properties of the QNM and branch-cut expansions and investigate their consistency with the excitation radius.

\begin{figure}[t]
\centering
\includegraphics[width=1\linewidth]{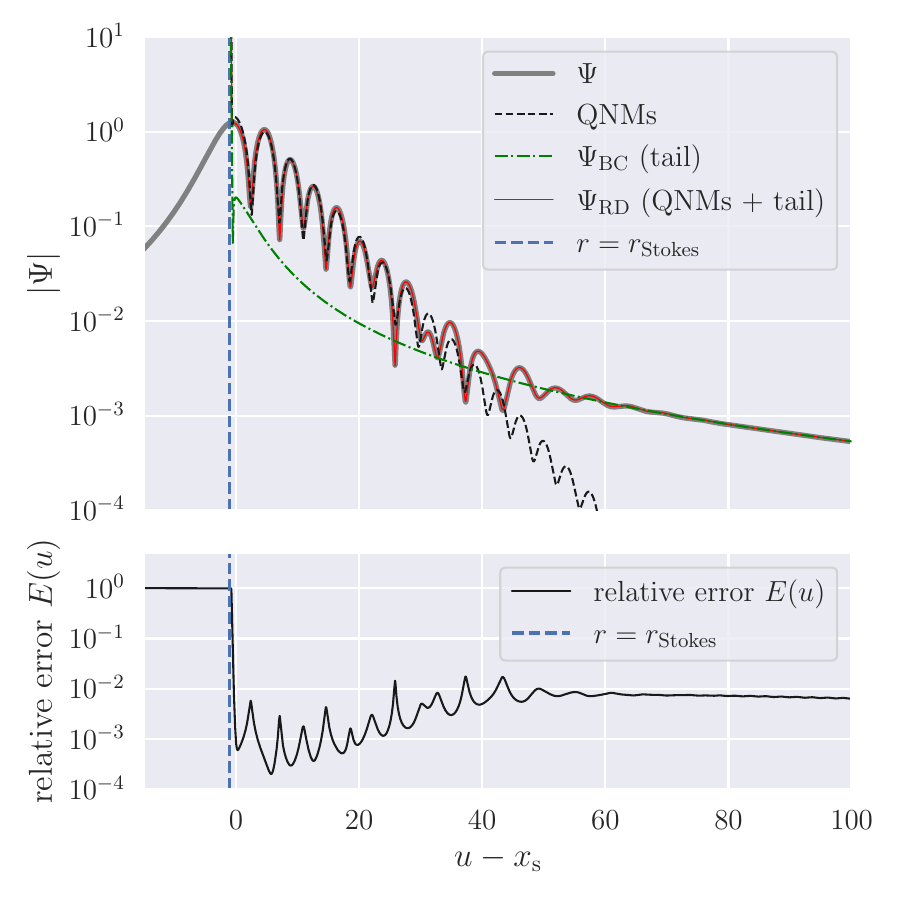}
\caption{Waveform reconstruction for the Kerr BH ($\chi =0.7$) and relative error between $\Psi$ and $\Psi_{\rm RD}$.
}
\label{fig_Sch_reconstruction}
\end{figure}

To this end, we compare the excitation radius obtained from the anti-Stokes structure with an independent estimate derived from the convergence properties of the QNM excitation factors. 
In this work, we compute the excitation factors or the branch-cut contribution associated with the Sasaki-Nakamura variable with the spin weight $s=-2$ \cite{Sasaki:1981sx}.
Given the in-mode homogeneous solution to the perturbation equation $\psi_{\rm IN} (x)$:
\begin{equation}
    \psi_{\rm IN} = 
    \begin{cases}
        &e^{-i k_{\rm H} x}\,,\\
        &A_{\rm out}(\omega) e^{+i \omega x} + A_{\rm in} (\omega) e^{-i \omega x}\,,
    \end{cases}
\end{equation}
where $k_{\rm H} \coloneqq \omega - m \Omega_{\rm H}$ and $\Omega_{\rm H} \coloneqq a/(2M r_+)$, the response from the perturbed BH with the source term $I(\omega)$ is \cite{Andersson:1996cm}
\begin{equation}
    \Psi = \int d \omega \frac{A_{\rm out} (\omega)}{A_{\rm in} (\omega)} \int dx' I(\omega,x') e^{-i \omega (t-x-x')}\,,
\end{equation}
where an observer is located at distant region, and the source term $I(\omega, x')$ is localized at $x'=x_{\rm s}$ with $2M \ll x_{\rm s} \ll x$.
For our purpose, i.e., identifying the site of QNM excitation, it is convenient to impose a localized source term of $I(\omega, x') = \delta (x'-x_{\rm s})$.
In this case, we have
\begin{equation}
\Psi = \int d \omega \frac{A_{\rm out} (\omega)}{A_{\rm in} (\omega)} e^{-i \omega (u-x_{\rm s})}\,,
\end{equation}
with $u \coloneqq t-x$, and the ringdown waveform is given by
\begin{equation}
\Psi_{\rm RD} = \sum_{\sigma,n} B_{n}^{(\sigma)} e^{-i \omega_n (u-x_{\rm s})}
+ \Psi_{\rm BC} (u)\,,
\end{equation}
where $\Psi_{\rm BC}$ is the contribution from the branch cut in $A_{\rm out} / A_{\rm in}$ \cite{Leaver:1986gd} (see also Appendix~\ref{app_methodology}), $\sigma = +$ or $-$ stands for the prograde or retrograde mode, respectively, and the excitation factor \cite{Leaver:1986gd} is defined as 
\begin{equation}
    B_{n} \coloneqq \frac{-2 \pi i A_{\rm out} (\omega_n)}{(d A_{\rm in} (\omega)/d\omega)_{\omega = \omega_n}}\,.
\end{equation}
At later times, say $u-x_{\rm s} > -2 x_{\rm con}$ with a constant $x_{\rm con}$, we see the agreement between $\Psi$ and $\Psi_{\rm RD}$ (FIG.~\ref{fig_Sch_reconstruction}).
As the delta-function source is initially located at $x=x_{\rm s}$ ($t=0$), it arrives at the site of QNM excitation $x = x_{\rm con}$ when $t = x_{\rm s} - x_{\rm con}$.
Then it is reflected back to the observer when $t = x + x_{\rm s} -2 x_{\rm con}$, i.e., 
\begin{equation}
u - x_{\rm s} = - 2 x_{\rm con}\,.
\end{equation}

\begin{figure}[t]
\centering
\includegraphics[width=1\linewidth]{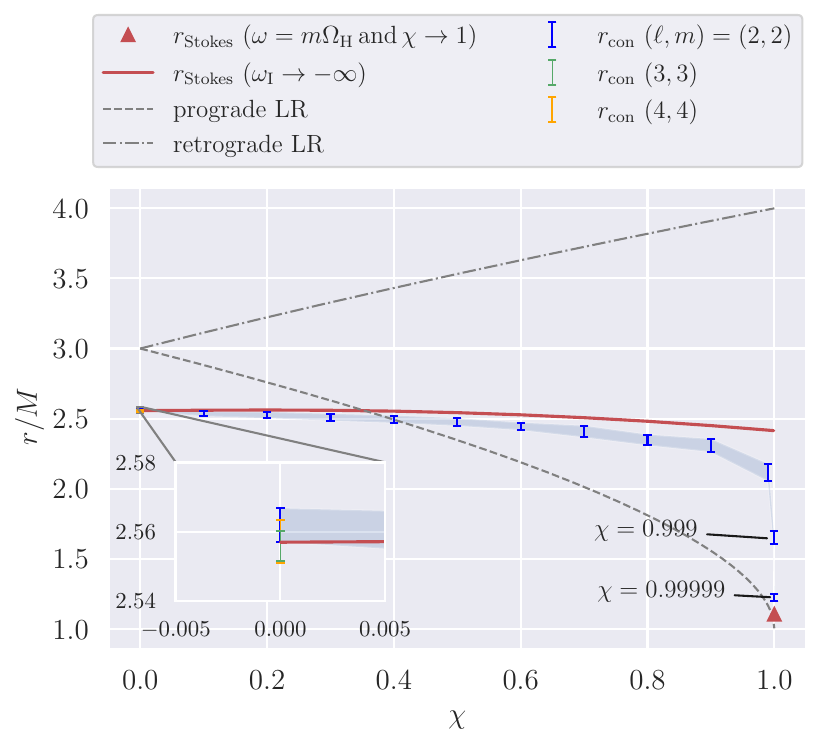}
\caption{Comparison between $r_{\rm con}$ (bars) and $r_{\rm Stokes}$ (red solid and red triangle).
The excitation radius $r_{\rm Stokes}$ is obtained in the high-overtone approximation, $\omega_{\rm I} \to - \infty$ (red solid) and in the ZDM approximation for the extremal limit, $\chi \to 1$ (red triangle). 
The triangle marker indicates the innermost excitation radius.
The bars indicate the variation in $r_{\rm con}$, resulting from the choice of the threshold, $10^{-2} \leq \epsilon_{\rm th} \leq 0.5$.
We set $\ell = m =2$ (blue bars) for a broad range of BH spins and additionally set $\ell = m =3$ (green) and $4$ (yellow) for the Schwarzschild case (see the inset plot).
We set $10^{-2}\leq \epsilon_{\rm
th} \leq 10^{-1}$ for $\chi = 0.99$ and $10^{-3}\leq \epsilon_{\rm
th} \leq 10^{-2}$ for $\chi=0.999$ and $0.99999$, since the smaller imaginary parts of the QNMs in the near-extremal regime make the deviation between the original waveform and the QNM+tail reconstruction more gradual, so the onset of the convergence breakdown is not sharply defined. 
We therefore adopt a smaller tolerance for identifying the convergence radius.
Prograde and retrograde light-ring radii (dashed and dot-dashed, respectively) are shown as reference.
}
\label{fig_Kerr_comp}
\end{figure}
The radius at which the QNM and branch-cut expansion ceases to converge, $x=x_{\rm con}$, is expected to matches with the excitation radius $r_{\rm Stokes}$. 
If our physical interpretation of the anti-Stokes line as the origin of QNM excitation is correct, the two estimates should agree, i.e., $r_{\rm Stokes} = r_{\rm con}$, where $r_{\rm con}$ is defined by $x(r_{\rm con}) = x_{\rm con}$.
FIG.~\ref{fig_Kerr_comp} summarizes the central result of this work by comparing
the excitation radius, $r_{\rm Stokes}$ (red solid line), predicted from the
Stokes geometry, with the convergence radius, $r_{\rm con}$ (bars), obtained from the ringdown reconstruction.
We find remarkable agreement between the two radii, especially for slowly and moderately rotating BHs, including the Schwarzschild case (FIG.~\ref{fig_Kerr_comp}).
We find that the excitation radius is located below the prograde light-ring radius for $\chi \lesssim 0.4$, while for $\chi \gtrsim 0.4$ it is located between the retrograde and prograde light-ring radii.
In the near-extremal limit, the value of $r_{\rm con}$ rapidly decreases towards the prograde light-ring radius $r_{{\rm LR}+}$ or the outer horizon.

To estimate the value of $r_{\rm con}$ (blue bars in FIG.~\ref{fig_Kerr_comp}), we numerically compute $\Psi_{\rm RD}$ and reconstruct the full waveform $\Psi$.
We then numerically find the value $x_{\rm con}$, defined by the condition that $\Psi_{\rm RD}$ ceases to converge to $\Psi$ at $u-x_0 = -2 x_{\rm con}$ (see FIG.~\ref{fig_Sch_reconstruction}).
To make the numerical ringdown reconstruction well defined, we adopt the following criterion: 
the reconstruction is deemed successful when the relative error of $\Psi$ and $\Psi_{\rm RD}$, $E(u)$, falls below a given threshold $\epsilon_{\rm th}$:
\begin{equation}
E(u) \coloneqq \left|\frac{\Psi(u) - \Psi_{\rm RD}(u)}{\Psi(u)} \right| < \epsilon_{\rm th}.
\end{equation}

FIG.~\ref{fig_Kerr_comp} shows the comparison between the QNM excitation radius $r_{\rm Stokes}$ and the convergence radius $r_{\rm con}$. 
The value of $r_{\rm con}$ for $\epsilon_{\rm th} = 10^{-2}$ is also tabulated in Table~\ref{tab:stokes_radius}.
We find better agreement for slowly and moderately rotating BHs. 

In the Schowarzschild case, we perform the ringdown reconstruction for higher angular modes: $\ell = 3$ and $4$.
We then find that even for the higher angular modes, the convergence radius $r_{\rm con}$ is stable at $r \simeq 2.55 \simeq r_{\rm Stokes} = 2.5569291$.
This is consistent with that the Stokes geometry and $r_{\rm Stokes}$ in the high-overtone limit has no dependence on $\ell$ [see Eq.~\eqref{V_q0}].
This is a very important observation, as it suggests that the QNM excitation radius does not coincide with the light ring at $r/M = 3$ even for higher angular modes, contrary to the conventional picture based on the eikonal limit \cite{Press:1971wr,1972ApJ...172L..95G,Ferrari:1984zz,Mashhoon:1985cya,Cardoso:2008bp,Yang:2012he}.

For rapidly rotating BHs, however, a noticeable discrepancy develops between $r_{\rm Stokes}$ and $r_{\rm con}$. 
The origin of this discrepancy can be understood from the approximations underlying the anti-Stokes analysis. 
The excitation radius $r_{\rm Stokes}$ (red-solid line in FIG.~\ref{fig_Kerr_comp}) is obtained from the Stokes geometry in the limit
\begin{equation}
    \omega_{\rm I} \rightarrow -\infty\,.
\end{equation}
However, rapidly rotating black holes have significantly smaller damping rates than slowly rotating ones.
In the extremal limit, the onset of QNM excitation is dominated by ZDMs \cite{Yang:2013uba}, whose frequencies do not belong to the highly damped sector captured by the asymptotic approximation $\omega_{\rm I} \to -\infty$.
Consequently, the Stokes geometry in the high-overtone approximation is not expected to accurately describe the long-lived QNMs that dominate ringdown signals for rapid spins.
In the following, we then identify the excitation radius from the anti-Stokes lines for ZDMs $\omega \to m \Omega_{\rm H}$, rather than $\omega_{\rm I} \to -\infty$.

We can simplify the Teukolsky equation by imposing $\omega \to m \Omega_{\rm H}$ for $\chi \to 1$ and by using the eikonal limit with $(a \omega/L)^2 \ll 1$ \cite{Yang:2013uba}, where $L \coloneqq \ell + 1/2$.
In the limit, the Teukolsky equation reduces to
\begin{equation}
    \left[ \frac{d^2}{dr^2} + V_{\rm ext}(r) \right] \psi(r) =0\,.
\end{equation}
where
\begin{align}
    V_{\rm ext} (r) \coloneqq \frac{U}{{\cal F}^2} -\frac{1}{2} \left( \log \, {\cal F} \right)'' -\frac{1}{4} \left[ (\log \, {\cal F})' \right]^2\,,
    \label{extremal_pot}
\end{align}
$U(r) \coloneqq (K^2 -\Delta \lambda)/(r^2+a^2)^2$, ${\cal F} \coloneqq \Delta/(r^2 + a^2)$, $K \coloneqq am - \omega (r^2+a^2)$, $\lambda \coloneqq a^2 \omega^2 - 2am \omega +A_{\ell m \omega}$, and the eigenvalue of the separation constant $A_{\ell m \omega}$ is
$A_{\ell m \omega} \simeq L^2 \left[ 1- \frac{a^2 \omega^2}{2 L^2} (1-\mu^2) \right]$ with $\mu \coloneqq m/L$, validated when $(a \omega/L)^2 \ll 1$ \cite{Yang:2013uba}.

\begin{figure}[t]
\centering
\includegraphics[width=0.48\textwidth]{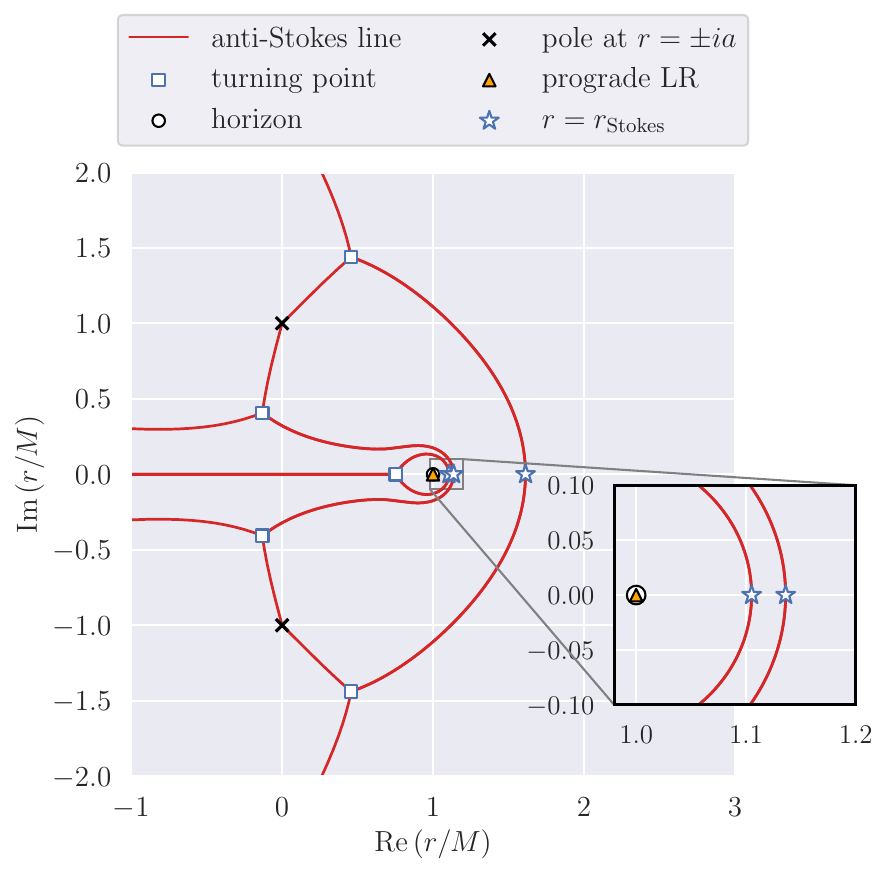}
\caption{
Stokes geometry of the Teukolsky equation with $\chi \to 1$ with the eikonal and ZDM approximations ($\omega = m \Omega_{\rm H}$).
For clarity, the Stokes lines are omitted from the plot.
}
\label{fig_stokes_099999}
\end{figure}

Even in the case of $\ell =2$, ZDM frequency in the extremal limit satisfies $(a \omega/L)^2 \simeq 0.16$, and the eikonal approximation gives a relatively accurate result~\cite{Yang:2013uba}.
The approximated potential barrier \eqref{extremal_pot} with $\chi \to 1$ has three anti-Stokes lines that intersect the real $r$-axis outside the horizon at $r/M = 1.10516429$, $1.136147450$ and $1.612406355$ (see FIG.~\ref{fig_stokes_099999}).
This observation would indicate that, in the extremal regime, the excitation of the ZDMs may not be characterized by a single excitation radius.
Instead, the Stokes geometry suggests multiple QNM excitation radii.
Since the complete QNM response can be established once the perturbation reaches the innermost excitation radius of QNM, it is natural to regard this radius as the effective QNM excitation radius: $r_{\rm Stokes} (\chi \to 1) = 1.10516429$ (red triangle in FIG.~\ref{fig_Kerr_comp}).
Indeed, this innermost excitation radius lies very close to the horizon in the extremal limit, consistent with the behavior of $r_{\rm con}$, which also approaches the horizon in the same limit (see FIG.~\ref{fig_Kerr_comp}).

Finally, we summarize several possible reasons why $r_{\rm con}$ and $r_{\rm Stokes}$ do not exactly agree, especially at intermediate or rapid BH spins.
A detailed investigation of these possible sources of discrepancy is left for future work.
{\bf (i)} The excitation radius depends on the QNM overtone number, as discussed above and in Appendix~\ref{app_spiral}, implying that the notion of a QNM excitation radius is not unique.
{\bf (ii)} Our estimate of $r_{\rm Stokes}$ is based on the Stokes geometry evaluated within either the high-damping approximation or the ZDM approximation in the eikonal limit, which may introduce quantitative deviations from the exact value.
{\bf (iii)} The anti-Stokes lines can intersect the real $r$-axis at multiple locations in the near-extremal cases (see FIG.~\ref{fig_stokes_099999}), which also imply the non-uniqueness of QNM excitation radius.

\section{Conclusion}
\label{sec_conclusion}

In this work, we investigated the excitation region of Kerr black hole (BH) quasinormal modes (QNMs). 
We proposed that the intersection of the anti-Stokes line of the Teukolsky equation with the real radial axis defines the QNM excitation radius. Since crossing the anti-Stokes line switches the dominance between the incoming and outgoing WKB solutions, we interpret the intersection as the origin from which QNMs dissipate toward both the horizon and spatial infinity.
This picture was examined from two complementary perspectives: an analytic analysis based on the Stokes geometry, and numerical calculations based on the ringdown reconstruction, including both QNM and branch-cut contributions. 
We found that these independent approaches yield consistent estimates of the excitation region of QNMs. 
Although the high-damping ($\omega_{\rm I} \to - \infty$) approximation becomes inadequate in the near-extremal regime, the Stokes geometry constructed for zero-damping modes is consistent with the QNM convergence radius defined by the ringdown reconstruction.

The identification of the QNM excitation radius has important implications for the long-standing problem of determining the onset of ringdown. 
In particular, once an excitation radius is specified in Kerr spacetime, the ringdown starting time can be discussed in a more direct connection with the dynamics of the merger process. 
Although the situation remains nontrivial for comparable-mass mergers, the interpretation is especially intriguing in the extreme-mass-ratio limit, where significant QNM excitation may occur as the companion object passes through the excitation radius.
In this sense, our results provide a new perspective on the physical origin of ringdown and may help bridge the gap between the plunge dynamics and the onset of linear QNM excitation. 
Further applications of this idea are left for future work.

Our analytic analysis relied on approximations in the determination of the Stokes geometry. 
A more precise treatment of the complex potential, performed numerically without relying on the approximations we made, may allow a more accurate determination of the QNM excitation radius and may reveal a nontrivial frequency dependence of the excitation radius.

The near-extremal regime deserves particular attention. 
In this limit, the excitation radii inferred from ringdown reconstruction deviate significantly from the predictions based on the approximated Stokes geometry with $\omega_{\rm I} \to - \infty$. 
This discrepancy may indicate that different overtones are excited at different effective radii, rather than sharing a single universal excitation radius among all overtones.
Motivated by this interpretation, we instead consider the Stokes geometry associated with the zero-damping modes, which may dominate ringdown signals, in the near-extremal cases.
The resulting innermost excitation radius is remarkably consistent with the trend of the numerical results of $r_{\rm con}$ in the near-extremal cases.
This suggests that for rapidly spinning black holes, the QNM excitation radius is governed by the long-lived QNMs that dominate the early ringdown, rather than by the asymptotic highly damped modes or retrograde ones.
Furthermore, the difference between the excitation radii obtained in the $\omega_{\rm I}\to -\infty$ limit and for $\omega=m\Omega_{\rm H}$ (corresponding to the ZDMs in the extremal limit) suggests that the excitation radius depends on the QNM overtone number.
More generally, the Stokes geometry depends on the complex
QNM frequency. 
Consequently, the excitation radius is
expected to depend on the overtone number.

The location at which QNMs become excited is far from obvious a priori since characteristic radii associated with Kerr black holes---such as the prograde and retrograde light-ring radii on the equatorial plane---are generally distinct. 
Despite the complexity of Kerr spacetime, the present work demonstrates that the Stokes geometry can be a powerful tool for identifying the physical origin of QNM excitation.
%
%

%%%%%%%%%%%%%%%%%%%%%%%%%%%%%%%%%%%%%%%%%%%%%%%%%%%%%%%%%
\acknowledgments
%%%%%%%%%%%%%%%%%%%%%%%%%%%%%%%%%%%%%%%%%%%%%%%%%%%%%%%%%
%
The author thank Vitor Cardoso, Yanbei Chen, Max Isi, Adrien Kuntz, Taiga Miaychi, Ryo Namba, Jaime Redondo-Yuste and Subhodeep Sarkar for fruitful discussions.
N.~O. was supported by Japan Society for the Promotion of Science (JSPS) KAKENHI Grant No.~JP26K17142.

\appendix 

\section{Spiral structure of anti-Stokes lines}
\label{app_spiral}

In Sec.~\ref{sec_stokes_geo}, we discussed the Stokes geometry in the high-overtone
limit by taking
\begin{equation}
\frac{\omega_{\rm R}}{\omega_{\rm I}}\to 0\,,
\end{equation}
for which the Stokes geometry takes the simple configuration (FIG.~\ref{fig_stokes_0000_0700}).
Here we discuss the configuration of anti-Stokes lines at finite complex QNM frequency with
\begin{equation}
0<|\omega_{\rm R}|\ll |\omega_{\rm I}|\,.
\end{equation}

In the Schwarzschild case, for example, the WKB phase contains the integral
\begin{equation}
z(r)=\int^r_0 V(r')\,dr'
    = r+\log(r-1) + i \pi,
\label{app_eqq_z_formula}
\end{equation}
where the lower endpoint is set at the turning point $r=0$, from which the relevant anti-Stokes line emanates. 
Because of the logarithmic term in \eqref{app_eqq_z_formula},
$z(r)$ is multi-valued. 
On the $p$-th Riemann sheet we may write
\begin{equation}
z_p(r) \coloneqq r+\log(r-1)+(2p + 1)\pi i\,, \ \ \ p\in \mathbb{Z}\,.
\end{equation}
\begin{figure*}[t]
\centering
\includegraphics[width=0.45\linewidth]{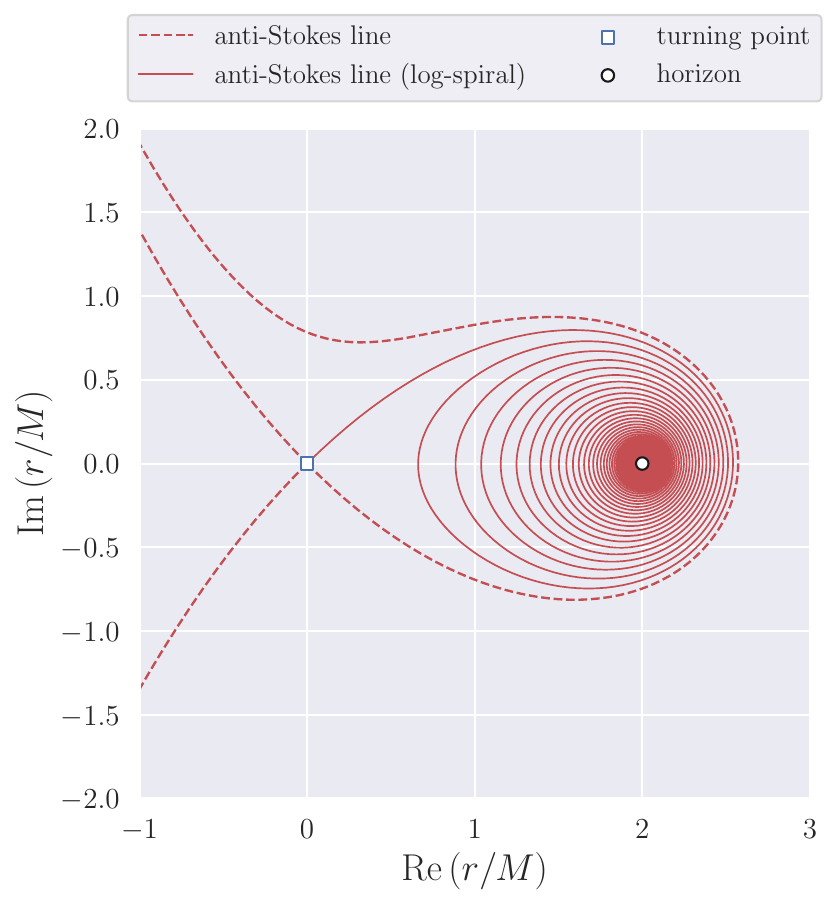}
\includegraphics[width=0.5\linewidth]{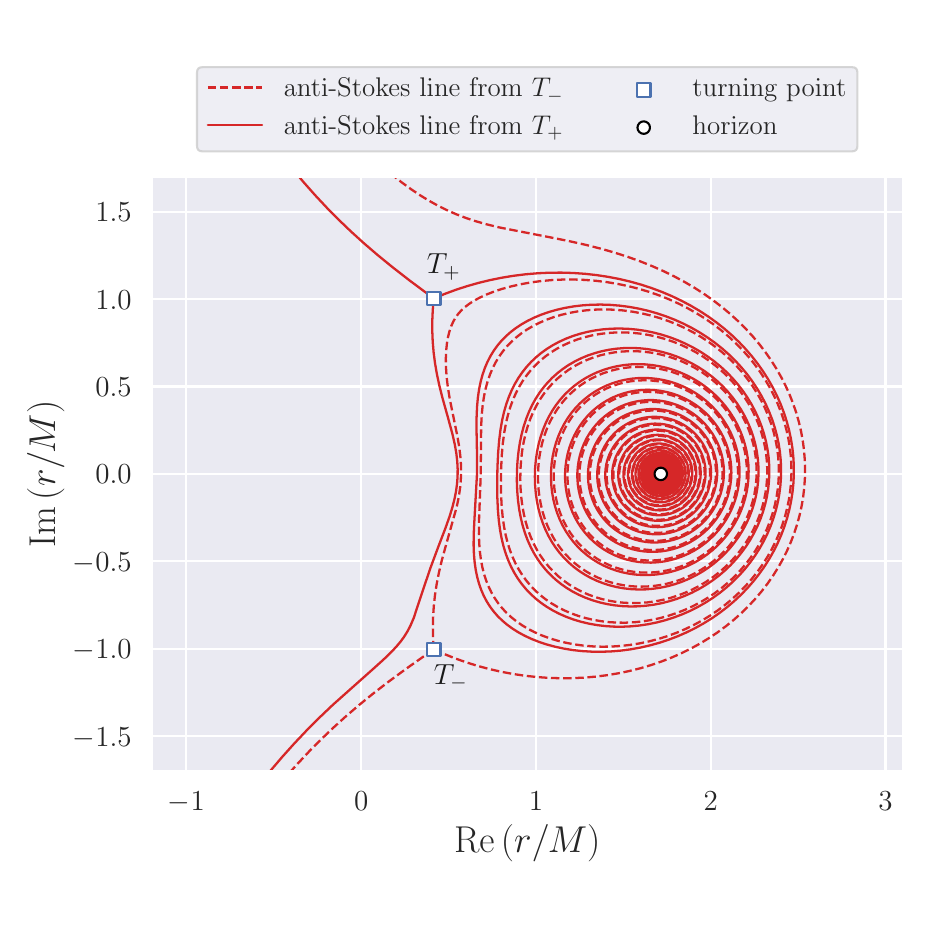}
\caption{
{\bf (Upper)}\, Anti-Stokes lines for the Schwarzschild case with the prograde QNM $(\ell,m,n) = (2,2,30)$.
{\bf (Lower)}\, Anti-Stokes lines for the Kerr case ($\chi = 0.7$) with the prograde QNM $(\ell,m,n) = (2,2,100)$.
}
\label{app_fig_anti_stokes_spiral}
\end{figure*}

Let us introduce new coordinates as $\rho e^{i\theta} = r-1$ with $\rho \in {\mathbb R}$ and $\theta \in {\mathbb R}$.
Then, we have
\begin{equation}
z_p
=
1+\rho e^{i\theta}
+\log\rho
+i(\theta+2\pi p).
\end{equation}
Writing \(\omega=\omega_{\rm R}+i\omega_{\rm I}\), the anti-Stokes
condition
\begin{equation}
{\rm Re} (i \omega z_p)=0
\end{equation}
becomes
\begin{equation}
\omega_{\rm I}
\left(1+\rho\cos\theta+\log\rho\right)
+
\omega_{\rm R}
\left(\rho\sin\theta+\theta+2\pi p\right)
=0 .
\label{app_antistokes_cond}
\end{equation}
For finite \(\omega_{\rm R}\), this equation describes logarithmic
spiraling of the anti-Stokes lines near the horizon. In particular, the
term $2\pi p$ is absent in the strict
$\omega_{\rm R}/\omega_{\rm I}\to0$ limit, but becomes relevant at
finite frequency.

On the real axis outside the horizon, i.e., $\theta=0$, the condition \eqref{app_antistokes_cond} is satisfied at $r = r_p$ where
\begin{equation}
r_p+\log(r_p -1)=-\frac{2\pi p\,\omega_{\rm R}}{\omega_{\rm I}}\,.
\end{equation}
Consequently, the same anti-Stokes line intersects the real-$r$ axis repeatedly at the sequence of points $r=r_p$ (FIG.~\ref{app_fig_anti_stokes_spiral}). These multiple intersections should not be interpreted as distinct QNM excitation radii. Rather, they are a consequence of the logarithmic monodromy of the WKB phase around the horizon.
Thus the multiple real-axis intersections at finite frequency are branch-dependent images of the same underlying Stokes structure. 
Normalizing the WKB solution on the branch at the far zone ($r \to \infty$), the physically relevant dominance switching is associated with the outermost intersection. 
In the high-overtone limit, on the other hand, all these branch-dependent intersections coalesce, since
\begin{equation}
\lim_{\omega_{\rm I} \to - \infty}-\frac{2\pi p\,\omega_{\rm R}}{\omega_{\rm I}} = 0\,.
\end{equation}
Hence, in the limit, the excitation radius for the Schwarzchild BH is uniquely determined as the root of 
\begin{equation}
r+\log(r -1)=0\,.
\end{equation}

\section{Ringdown reconstruction with Kerr QNMs and branch-cut tail}
\label{app_methodology}
In our ringdown reconstruction from QNMs and branch-cut contribution, prograde and retrograde QNMs are included up to $n = N_+$ and $n = N_-$, respectively, and the list of $N_+$ and $N_-$ for each BH spin is summarized in Table~\ref{tab:qnm_numbers}.
For the Schwarzschild case with $\ell =2$, we omit the 8th and 9th overtones from the QNM sector as their contribution is included in $\Psi_{\rm BC}$ in our computation.
For $\ell = 3$ and $4$ in the Schwarzschild case, we included prograde and retrograde modes up to $N_+ = N_- = 89$ and $N_+ = N_- = 113$, respectively.
\begin{table}[b]
\centering
\begin{tabular}{ccc}
\toprule
 $\chi$ & $N_+$ & $N_-$ \\
\midrule
0 & 99 & 99\\
0.1 & 99& 99\\
0.2 & 99& 99\\
0.3 & 99& 99\\
0.4 & 99& 99\\
0.5 & 99& 99\\
0.6 & 99& 49\\
0.7 & 99& 49\\
0.8 & 99& 39\\
0.9 & 99& 33\\
0.99 & 171& 19\\
0.999 & 173& 14\\
0.99999 & 435 & 10\\
\bottomrule
\end{tabular}
\caption{Number of prograde and retrograde QNMs included in the ringdown reconstruction for each spin with $\ell = m=2$.}
\label{tab:qnm_numbers}
\end{table}
In this paper, the numbers of prograde and retrograde QNMs, $N_{\pm}$, adopted in the ringdown reconstruction are determined by requiring that the convergence time of the QNM-plus-tail reconstruction $\Psi_{\rm RD}$ to the original waveform $\Psi$ changes by less than $1\%$ when the maximum overtone number $N_{\pm}$ is reduced to $N_{\pm}-10$.
We compute QNM frequencies and the eigenvalue of the separation constant by using Leaver's method \cite{Leaver:1985ax,Leaver:1986gd,Leaver:1986vnb}.
The excitation factors are computed by using the MST method \cite{Mano:1996vt}.

The contribution from the branch cut in the Green's function is also computed by the MST method.
We compute the value of $A_{\rm out} (\omega) / A_{\rm in} (\omega)$ in the Sasaki-Nakamura variable at the both side of the branch cut, i.e., at $\omega = \pm \delta_{\rm R}+i \omega_{\rm I}$ with $\delta_{\rm R} = 10^{-5}$ and $\omega_{\rm min} \leq  \omega_{\rm I} \leq 0$.
The value of $\omega_{\rm min}$ is chosen such that the convergence radius $r_{\rm con}$ changes by less than $1\%$ when the cutoff is shifted from $M\omega_{\rm min}$ to $M\omega_{\rm min}-5$.
We find that $|M\omega_{\rm min}|=15$ is sufficiently large to satisfy this criterion.
Then the branch-cut contribution $\Psi_{\rm BC} (u)$ is obtained as
\begin{align}
\begin{split}
   \Psi_{\rm BC} (u) &= \int_{\delta_{\rm R} +i \omega_{\rm min}}^{\delta_{\rm R}} d\omega  \frac{A_{\rm out} (\omega)}{A_{\rm in} (\omega)} e^{-i \omega (u-x_{\rm s})}\\
   &- 
   \int_{-\delta_{\rm R} +i \omega_{\rm min}}^{-\delta_{\rm R}} d\omega  \frac{A_{\rm out} (\omega)}{A_{\rm in} (\omega)} e^{-i \omega (u-x_{\rm s})}\,.
   \end{split}
\end{align}

\bibliography{references}

\end{document}